\documentclass[11pt]{article}
\pdfoutput=1
\usepackage{mathpazo}
\usepackage{color}
\usepackage{etoolbox}
\newbool{submit}
\setbool{submit}{true}
\newbool{havealerts}
\newbool{fullversion}
\setbool{fullversion}{true}

\usepackage{nicefrac}

\newcommand{\nfrac}{\nicefrac}

\usepackage[compact]{titlesec}
\usepackage{mdwlist}

\usepackage{geometry, pdfsync, tabularx,graphicx}
\usepackage{amsmath, amsthm, amssymb, enumerate}
\usepackage{marginnote, fullpage}

\ifbool{submit}{\renewcommand{\marginnote}[1]{}}{}
\ifbool{submit}{\setbool{havealerts}{false}}{\setbool{havealerts}{true}}
\newtheorem{theorem}{Theorem}[section]
\newtheorem{claim}[theorem]{Claim}

\newtheorem{definition}[theorem]{Definition}

\newtheorem{lemma}[theorem]{Lemma}

\newcommand{\defeq}{\stackrel{\textup{def}}{=}}
\newcommand{\eps}{\varepsilon}

\newcommand{\alert}[1]{\ifbool{havealerts}{\marginnote{\textcolor{red}{#1}}}{}}

\newcommand{\F}[0]{\ensuremath{\mathcal{F}}}

\newcommand{\Z}[0]{\ensuremath{\mathbb{Z}}}

\renewcommand{\epsilon}[0]{\varepsilon}

\newcommand{\shrinkspace}[1]{\ifbool{fullversion}{}{\vspace{#1}}}

\renewcommand{\F}[0]{\ensuremath{\mathbb{F}}}

\newcommand{\fq}{{\sc $\Fq$-QCSPP }}

\newcommand{\q}{{{\F_q}}}
\renewcommand{\to}{\mapsto}

\renewcommand{\[}{\left[}
\renewcommand{\] }{\right]}

\newcommand{\ml}{{\mathcal{L}}} 
 
\newcommand{\Fq}{{\mathbb{F}_q}}

\numberwithin{equation}{section}

\newtheorem{Lemma}[theorem]{Lemma}

\newtheorem{Fact}[theorem]{Fact}

\newtheorem{Alg}{Algorithm}

\def\bull{\vrule height .9ex width .8ex depth -.1ex }

\title{\bf $2^{\log^{1-\eps} n}$ Hardness for Closest Vector Problem with Preprocessing}  
\author{Subhash A. Khot  \\  
 {\small New York University} \\
{\small New York, NY, USA.}\\
{\small  khot@cims.nyu.edu}
 \and Preyas Popat 
   \\ 
 {\small New York University} \\
{\small New York, NY, USA.}\\
{\small  popat@cs.nyu.edu}
 \and Nisheeth K. Vishnoi
     \\ {\small Microsoft Research}  \\
{\small Bangalore, India}\\
{\small nisheeth.vishnoi@gmail.com}}
\date{}
\begin{document}

\maketitle



\begin{abstract}
We prove that for an arbitrarily small constant  $\eps>0,$ assuming NP$\not \subseteq$DTIME$(2^{{\log^{O(1/\eps)} n}})$, the preprocessing versions of the closest vector problem and the nearest codeword problem are hard to approximate within a factor better than $2^{\log ^{1-\eps}n}.$ This improves upon the previous hardness factor of $(\log n)^\delta$ for some $\delta > 0$ due to \cite{AKKV05}. 

\end{abstract}

\section{Introduction}

Given an integer lattice $B$ and a target vector $t$ in $\Z^m$, the {\sc Closest Vector Problem (CVP)} asks for the vector in $B$ nearest to $t$ under the  $l_p$ norm.  All $p \geq 1$ are interesting although, $p=2$ case has received the most attention.  An integer lattice is a set of vectors $\{ \sum_{i=1}^n \alpha_i { b}_i \ | \ \alpha_i \in \Z \}$, where ${b}_1, { b}_2, \ldots, { b}_n \in \Z^m$ is a set of linearly independent vectors, called the \emph{basis} of the lattice.  An important variation of {\sc CVP} is the pre-processing version of the problem where the lattice $B$ is known in advance and the algorithm is allowed arbitrary pre-processing on $B$ before the input $t$ is revealed.   This is known as the {\sc Closest Vector Problem with Pre-processing (CVPP)}.  A related problem is the {\sc Nearest Codeword Problem} ({\sc NCP}) where the input is a generator matrix $C$ of a linear code over $\F_{2}$ and a target vector $t$.  The goal is to find the codeword nearest to $t$ in Hamming distance. Again, if $C$ is known in advance and arbitrary pre-computation is allowed on it, the problem is known as the {\sc Nearest Codeword Problem with Pre-processing} ({\sc NCPP}).

Pre-processing problems
arise in cryptography and coding theory where,  typically, a
publicly known lattice (or a linear error-correcting code) is being used to
transmit messages across a faulty channel. The
decrypting or decoding of the received word is equivalent  to solving an instance of {\sc CVP}  for the
lattice being used in the protocol.  The basis of the lattice being
 known beforehand, it becomes imperative to understand if the performance of the
decoding algorithm can be improved, or if the security of the cryptographic
protocol can be compromised (see \cite{cvpp-FM,Regev04} for more details).

Potentially, this pre-computated information  could make CVPP easier than {\sc CVP}. 
Indeed,  using the so-called Korkine-Zolotarev
basis, Lagarias {\em et al.}  \cite{LLS90} constructed an $O(n^{1.5})$
factor approximation algorithm for {\sc CVPP}, which is significantly better than the best known 
almost-exponential $2^{O(n \log \log n / \log n)}$ approximation factor known for {\sc CVP}, see \cite{MicciancioV10,Schnorr87}.  This $n^{1.5}$ factor was
 improved to $n$ by Regev \cite{Regev04}, and subsequently
to $O(\sqrt{n/\log n})$ by Aharonov and Regev \cite{AR05}.

As for the inapproximability of {\sc CVP},  it was proved by Dinur {\em et al.} \cite{DKRS03} that it is  NP-hard to approximate to a factor within $n^{c/\log \log n}$ for some constant $c > 0$.   This improved an earlier result of \cite{ABSS97} showing that it is quasi-NP hard to approximate to a factor within $2^{{\log }^{1 - \epsilon} n}$ for any constant $\epsilon > 0$. 
Obtaining inapproximability results for {\sc CVPP} has been a more challenging task: 
Feige and Micciancio
\cite{cvpp-FM} proved a $\nfrac{5}{3} -\epsilon$ factor NP-hardness for
{\sc NCPP} for any constant $\epsilon > 0$.
 This was improved to $3-\epsilon$ by Regev
\cite{Regev04}. These authors observed that
a factor $C$ hardness for {NCPP} implies a factor $C^{1/p}$ hardness
for {CVPP} under the $\ell_p$ norm for any $1 \leq p < \infty.$ 
Also, hardness results  in the $\ell_{\infty}$ case can be obtained by using the norm-embedding technique due to Regev and Rosen \cite{RegevR06}.

The inapproximability results were improved in \cite{AKKV05} who proved a factor $C$ NP-hardness for {\sc CVPP} and {\sc NCPP} for any constant $C.$ \cite{AKKV05} showed how their result can be extended to a hardness of $(\log n)^\delta$ for some $\delta > 0$ under the assumption that NP $\not\subseteq$ DTIME($2^{poly (\log n)}$).  They also give another reduction which achieves a hardness factor of $(\log n)^{1 - \epsilon}$ for {\sc NCPP} for any $\epsilon > 0$.  This latter reduction is under a certain unproved hypothesis about a pre-processing version of the PCP Theorem. \footnote{The authors claim to have a proof, but do not include it in the paper.}

\subsection{Main Result and Overview} 
\label{overview}

The following is the main theorem of this paper.
\begin{theorem}[Main Theorem]\label{thm:main-cvpp}
Unless  $NP \subseteq DTIME(2^{\log^{O(1/\eps)} n})$, {\sc NCPP} and {\sc CVPP} are hard to approximate to a factor within $2^{{\log^{1-\epsilon}} n}$ for an arbitrarily small constant $\epsilon > 0$.
\end{theorem}

 This  improves on the previous hardness factor of $(\log n)^\delta$ for some $\delta > 0$ due to \cite{AKKV05} and essentially matches the almost polynomial factor inapproximability of Dinur {\em et al.} \cite{DKRS03} for {\sc CVP}. 
 We emphasize that unlike the case of  {\sc CVP} where the best approximation algorithm achieves a factor of $2^{n\log\log n/\log n},$ the best approximation algorithm for {\sc CVPP} achieves an approximation factor of $O(\sqrt{n/\log n}).$ 
 
\paragraph{Overview of the proof.} 
We will show a hardness factor of $2^{{\log^{1- \epsilon} n}}$ for the {\sc Minimum Weight Solution Problem with Pre-processing (MWSPP)}.  The input to this problem consists of a set of \emph{fixed} linear forms described by ${B}_f \in \mathbb{F}_2^{l \times N}$ , a set of \emph{variable} linear forms ${ B}_v \in \mathbb{F}_2^{l' \times N}$ and a target vector ${ t} \in \mathbb{F}_2^l$.  The goal is to find a solution ${ x} \in \mathbb{F}_2^N$ to the system ${ B}_f { x} = { t}$, which minimizes the Hamming weight of the vector ${ B}_v { x}$.  We allow arbitrary pre-processing on all parts of the input except the vector ${ t}$.  It is easy to check that {\sc MWSPP} is a reformulation of {\sc NCPP}. Henceforth, we focus on the {\sc MWSPP} problem. See Section \ref{sec:prelim} for preliminaries and definitions and the equivalence of MWSPP with NCPP. 

Our reduction builds on the second reduction of \cite{AKKV05} to {\sc MWSPP}.  The authors in \cite{AKKV05} make a certain hypothesis about the pre-processing version of the PCP Theorem.  This hypothesis leads to the hardness of approximation of a pre-processing version of the {\sc Label Cover} problem.  Recall that an instance of {\sc Label Cover} is given by a bipartite graph $G = (V,W,E,[R], [S])$ and for each edge $e = (v,w) \in V \times W$, a function $\pi_e : [R] \mapsto [S]$.  A labeling to the graph consists of an assignment $A : V  \mapsto [R], \ W \to [S] $.  An edge $e = (v,w)$ is said to be satisfied by an assignment $A$ if $\pi_e(A(v)) = A(w)$.  The value of an instance is the maximum fraction of edges that can be satisfied by any labeling.

It is a consequence of the PCP Theorem \cite{AS98, ALMSS98} and Raz's Parallel Repetition Theorem \cite{Raz98} that for every constant $R$, given an instance of {\sc Label Cover} it is NP-hard to distinguish whether the value of the instance is $1$ or at most $R^{-\gamma}$ for some absolute constant $\gamma > 0$.  The authors in \cite{AKKV05} show that a similar hardness holds for the {\sc LCPP} problem under their hypothesis.  In the {\sc LCPP} problem, the label set $[R]$ for each vertex $v \in V$ comes with a partition, and an {\em allowable set} from the partition.  The vertices in $V$ are required to be assigned labels from their respective allowable sets.   Pre-processing is allowed  on all parts of the {\sc LCPP} instance except for (the choice of) the allowable set for each vertex $v \in V$.

The reduction of \cite{AKKV05} from {\sc LCPP} to {\sc MWSPP} uses constructions of {\sc Label Cover} with an additional property called \emph{smoothness}.  An instance of {\sc LCPP} is called $\delta$-smooth if any two labels $i \neq i'$ of $v \in V$ map to different labels of $w \in W$ with probability at least $1 - \delta$ over the choice of neighbors $w$ of $v$.  The smoothness property was introduced in \cite{Khot02b} and has been used for several hardness of approximation reductions \cite{FGRW09, GRSW10, KS11}.  The hardness factor achieved by the the reduction from {\sc LCPP} to {\sc MWSPP} is bounded by $1/\delta$ and $1/s$ where $\delta$ is the smoothness parameter and $s$ is the soundness of the {\sc LCPP} instance.  The reduction of \cite{AKKV05} fails to give a hardness factor better than $(\log n)^{1 - \epsilon}$ for {\sc MWSPP} (even assuming their hypothesis) because they use constructions of {\sc Label Cover} which require size $n^{\Omega(1/\delta)}$ to ensure $\delta$-smoothness.  To get a better hardness factor using this reduction, we require instances of {\sc LCPP} with {\em very good} smoothness and soundness simultaneously (relative to the size of the instance).

Our main technical contribution is to construct instances of {\sc Hyper-graph Label Cover with Pre-processing (HLCPP)} with {\em very good} soundness and smoothness.  We achieve this using the  low degree test \cite{AS03, RS97},  which is guaranteed to work even for very small success probability, and combine it with the sum check protocol \cite{LFKN92}, which is used to reduce the number of queries.  The {\sc HLCPP} problem we consider is a labeling problem similar to {\sc LCPP} which differs from the latter as follows.

\begin{itemize}

\item The vertex set is multi-layered. 

\item The constraints are given by hyper-edges rather than edges.  A hyper-edge contains several edges and the constraint associated with a hyper-edge is a boolean AND of constraints associated to all its edges.

\item The constraints associated to edges are not \emph{many-to-one} (projection) constraints as in {\sc LCPP} but the more general \emph{many-to-many} constraints.
\end{itemize}

We give an outline of our reduction and the analysis below.

\paragraph{Our reduction and analysis.}
We will start with an instance of $\F_q$-{\sc Quadratic Constraint Satisfaction Problem} ($\F_{q}$-QCSP) for $q=2^r$.  The instance consists of $k$ homogeneous degree $2$   polynomial equations over  $\F_{q}$ with $n$ variables, where $k={\rm poly}(n).$  Each equation is of the form $p(z_{1},\ldots,z_{n})=v,$ and further, depends on at most $3$ variables. It can be shown that deciding if there is an assignment which satisfies all the equation is {\sc NP}-hard (see Theorem \ref{np-complete}), even when the l.h.s. of these equations ($p$'s) are available for pre-processing. We denote the pre-processing version by $\F_{q}$-QCSPP. Our first step is to boost soundness, i.e., to reduce the fraction of satisfied equations by any assignment, while keeping the number of variables small. This is done by combining an instance of \fq with an appropriate  Reed-Muller code over $q$.  We will eventually set  $q= n^{\log^{O(1/\eps)} n}.$
This allows us to construct an $\F_{q}$-QCSPP instance where it is hard to distinguish between perfectly satisfiable instances and those where any  assignment satisfies at most $k/q$ fraction of the polynomial equations.   An important feature of this reduction is that the variable set remains the same, so the number of variables is $n$, number of equations is $q$ and the
soundness is $k/q$ (which is essentailly same as $1/q$). This quantitative
setting of parameters is crucial for our result as the number of variables becomes negligible compared to the number of equations, and the reciprocal of the soundness.  The details of this reduction appear in Section \ref{sec:boost}.

Each equation can now depend on almost all of the $n$ variables and the next task is to deal with this. This is done by reducing checking an assignment for such a system of polynomial equations to the  task to constructing a PCP which makes $O(\log n)$ queries  and has soundness $1/q^{e}$ for some small constant $e>0.$  This is achieved  by combining the low degree test and  
the sum check protocol and is the technical heart of the PCP construction. 

First, the variables are identified with $\{0,1\}^{\log n}$ and embedded as a subcube of $\F_{q}^m$ where $m \defeq \log n.$ With this mapping, any assignment can be thought of as a function from $\{0,1\}^{m}$ to $\F_q$ and can be encoded as a polynomial over $\F_{q}^m$ of degree at most $m.$  In this setting, if the equation was $\sum_{i,j \in [n]} c(i,j)z_{i}z_{j}=v = \sum_{\alpha,\beta \in \{0,1\}^{m}} c(\alpha,\beta)z({\alpha}) z({\beta});$ $z,c$ can be thought  of as  polynomials of degree at most $m$ and $2m$ respectively. 
The Arora-Sudan points-vs-lines low degree test can be employed to ensure that $z$ corresponds to a small list of degree $m$ polynomials (assignments).  This test is able to list-decode an assignment with success probability as low as $1/q^{e}$ for some small constant $e>0.$ 

Once an assignment for the variables can be decoded, the task of verifying the polynomial equations $\sum_{\alpha,\beta \in \{0,1\}^{m}} c(\alpha,\beta)z({\alpha}) z({\beta})=c$ is equivalent to performing a weighted sum check over the sub-cube $\{0,1\}^m$.   We use the sum-check protocol of \cite{LFKN92} to verify that the decoded assignment satisfies the equations.
It can be shown that the soundness of the combined low degree test and the sum check protocol is at most $1/q^{f}$ for a small constant $f>0.$

The result is a PCP with $2m+2=O(\log n)$ layers where the first $2m$ layers correspond to the sum check protocol while the last two layers correspond to the {\em lines}  and the {\em points} table respectively.  Only the values to be assigned to the first table by the prover will depend on the r.h.s. of the \fq instance. 
Further, the use of low degree polynomials in encoding the assignments implicitly  gives our PCP {\em smoothness} properties which are used in the final reduction.  
While the preliminaries of the low degree test and the sum check protocol appear in Sections \ref{sec:low-degree} and \ref{sec:sum-check} respectively, the PCP construction appears in Section \ref{sum-check}.

This view of the PCP naturally leads us to constructing an {\sc HLCPP} instance which is the starting point of the reduction to {\sc MWSPP} and appears in Section \ref{graph}. 
Finally, the reduction from {\sc HLCPP} to {\sc MWSPP} is similar to the reduction of \cite{AKKV05} from {\sc LCPP} to {\sc MWSPP}. This appears in Section \ref{reduction}.  For the reduction to work, we define a notion of smoothness for {\sc HLCPP} which is similar to the one for {\sc LCPP} and we also need that the hyper-edges of the graph satisfy a uniformity condition which is inherited from the PCP construction.

The main differences in our reduction compared to the reduction of \cite{AKKV05} are the following:

\begin{itemize}

\item As mentioned earlier, the constraints in the {\sc HLCPP} graph are many-to-many constraints rather than many-to-one constraints.  However, the earlier reduction to {\sc MWSPP} still goes through in a relatively straightforward manner.

\item We manage to construct an instance of {\sc HLCPP} where the smoothness and soundness are both at most $1/q^f$ for some absolute constant $f > 0$ and the size of the instance is $q^{O(m)}$.  Here $m = \log n$ where $n$ is the number of variables in the original \fq instance.  It is not clear that such constructions are possible if we stick to the {\sc LCPP} problem.  The hardness factor can be made essentially as large as $q^{1/m}$ and we set $q$ to be very large compared to $m$ to get a good hardness factor relative to the size of the instance.  Specifically, we set $q= n^{\log^{O(1/\eps)} n}.$

\end{itemize}

\section{Preliminaries} 
In this section we state the problems we will consider and state basic results which will be useful in the construction of our PCP and the reduction.

\subsection{Problem Definitions and Basic Results}\label{sec:prelim}
We first define the quadratic CSP problem and its pre-processing version that will be a starting point of our reduction.
\begin{definition} $\F_{q}$-{\sc Quadratic CSP ($\F_{q}$-QCSP):} A {\sc  $\F_{q}$-QCSP} instance $Q \defeq \left(\{p_j\}_{j=1}^m, \{c_j\}_{j=1}^m\right)$ consists of a set of polynomial constraints over  variables $\{z_1,z_2,\ldots,z_n\}$.  Each equation is of the form $$p_j(z_1,z_2,\ldots,z_n) =c_j,$$ where $p_j$ is a homogeneous polynomial of degree $2$, and $c_j \in \F_{q}$.  The goal is to find an assignment to the variables  $\{z_1,z_2,\ldots,z_n\}$ each taking a value in $\mathbb{F}_{q}$ which satisfies as many constraints as possible.  Let ${OPT}(Q)$ denote the maximum, over assignments to the variables of $Q$, of the fraction of equations satisfied.

\end{definition}

\begin{definition}  $\F_{q}$-{\sc Quadratic CSP with Pre-processing ($\F_{q}$-QCSPP):} Given a  $\F_{q}$-{\sc QCSP} instance 
$$Q \defeq \left(\{p_j\}_{j=1}^m, \{c_j\}_{j=1}^m\right)$$ over variables $\{z_1,z_2,\ldots,z_n\}$ taking values in $\mathbb{F}_q,$ the  $\F_{q}$-{\sc QCSPP} problem allows arbitrary pre-processing on the polynomials $\{p_j\}_{j=1}^m$ before the inputs $\{c_j\}_{j=1}^m$ are revealed. 
\end{definition}


%
%
%

The following theorem can be proved in a similar manner as Theorem 4.2 in  \cite{AKKV05}.  We include a proof in Section \ref{sec:np-complete}.

\begin{theorem} \label{np-complete}

{\sc $\F_q$-QCSPP} is NP-complete for all $q=2^r$.

\end{theorem}

Next we define the problem that we prove is hard to approximate and show that it is equivalent to the nearest codeword problem with pre-processing.    
\begin{definition} {\sc Minimum Weight Solution Problem with Pre-processing (MWSPP):} \label{MW}
The input to this problem consists of a set of \emph{fixed} linear forms described by ${B}_f \in \mathbb{F}_2^{l \times N}$ , a set of \emph{variable} linear forms ${B}_v \in \mathbb{F}_2^{l' \times N}$ and a target vector ${ t} \in \mathbb{F}_2^l$.  The goal is to find a solution ${ x} \in \mathbb{F}_2^N$ to the system ${ B}_f { x} = { t}$, which minimizes the Hamming weight of the vector ${ B}_v { x}$.  We allow arbitrary pre-processing on all parts of the input except the vector ${ t}$.

\end{definition}

\begin{definition}{\sc Nearest Codeword Problem with and without Pre-processing:}
    An instance of {\sc NCP} is
    denoted by $(C,t)$ where $C \in \F_{2}^{n \times k},$ $t \in \F_{2}^{n}.$
  The goal is to find a solution ${ x} \in \mathbb{F}_2^k$ which minimizes the Hamming distance between $C { x} $ and  ${ t}.$ In the pre-processing version, {\sc NCPP}, we allow arbitrary pre-processing on all parts of the input except the vector ${ t}$.
\end{definition}

We note that {\sc MWSP} is actually same as the {\sc NCP} problem in disguise, though
we find it convenient to think of it as a separate problem.
To see the equivalence with
{\sc NCP}, let $x_0$ be a fixed vector such that ${ B}_{f} x_0 = t$, let
$w = { B}_{v}  x_0$  and
consider the code ${C} \defeq \{ { B}_{v} x \ | \ \text{ $x$ s.t. } { B}_{f}  x = { 0} \}$. Then
$$  \min_{x : { B}_{f}  x = { 0}}
 \delta(w, { B}_{v}  x ) =  \min_{x : { B}_{f}  (x+x_0) = t }
\delta({ B}_{v}  x_0 , { B}_{v}  x ) =
\min_{ x : { B}_{f}  x = t }  { wt} ( { B}_{v}  x ). $$
Here $\delta(\cdot,\cdot)$ measures the Hamming distance and $wt(\cdot)$ denotes the Hamming weight of a string.

Finally we note that proving the hardness for NCPP implies the hardness for CVPP.

\begin{theorem} \cite{cvpp-FM} \label{cvpp}
Let $1\leq p < \infty.$ If {\sc NCPP} ({\sc MWSPP}) is hard to approximate to factor $f$ then {\sc CVPP}, under the $l_p$ norm, is hard to approximate to factor $f^{1/p}$.
\end{theorem}

\subsection{Boosting Soundness through Codes}\label{sec:boost}

The following lemma shows how to boost soundness of the \fq instance although it increases the number of variables per equation. The proof of this lemma employs Reed-Muller codes and appears in Section \ref{sec:boost-proof}. 
\begin{Lemma} \label{sound}
Let $Q$ be an instance of \fq over $n$ variables and $k = poly(n)$ equations, for any $q=2^r$.  There is an instance $P$ of \fq over the same set of variables and $q$ equations such that:
\begin{itemize}
\item If $OPT(Q) =1$ then $OPT(P) = 1$ and 
\item if $OPT(Q) < 1$ then $OPT(P) \leq k/q.$
\end{itemize}
\end{Lemma}
In our reduction $q$ would be $n^{\log^{O(1/\eps)} n}$ and, hence, $q \gg k.$

%
%
%
%
%

\subsection{Low Degree Test}\label{sec:low-degree}
Now we move on to developing tools necessary for keeping the number of queries in our PCP small. The first step in this is the Low Degree Test. In this section we recall the basics, the test  and state the Arora-Sudan theorem which will be used.

An affine line in $\F_{q}^{m}$ is parametrized by $(a,b) \in (\F_{q}^{m} \backslash \{0\}) \times \F_{q}^{m}$ such that $ L_{a,b} \defeq \{ ax + b: x \in \F_{q}\}.$ Sometimes, we will drop the subscript if it is clear from the context. In what follows, if it helps, one can think of $m \defeq \log n$ and $d\defeq m$ as will be the case in our reduction.  For a polynomial $g: \F_{q}^{m} \mapsto \F_{q}$ of degree $d$ and a line $L\defeq L_{a,b},$ let $g|_{L}$ be the restriction of $g$ defined as $g|_{L}(x) \defeq g(ax+b)$  for $x \in \F_{q}.$ For two polynomials $g,h$ we denote $g \equiv h$ if they are identical.

\begin{definition}[{\bf Low Degree Test}] \label{low-degree-def}
The Low Degree Test takes as input the value table of a  function $f : \F_q^m \mapsto \Fq$ and for every (affine) line $L$ of $\mathbb{F}_q^m$, the coefficients of a degree $d$ polynomial $g_L$.  

The goal is to check that $f$ is a degree $d$ polynomial.  The intention is that $g_L$ is the restriction of $f$ to the line $L$.

The test proceeds as follows:

\begin{enumerate}

\item Pick a random point $x \in \F_q^m$ and a random line $L$ containing $x.$

\item Test that $g_L(x) = f(x).$

\end{enumerate}

\end{definition}

The following theorem can be inferred from Theorem 1 and Lemma 14 in \cite{AS03}.

\begin{theorem}[{\bf Soundness of Low Degree Test}]  \label{as}
There are absolute constants $0 < c_1, c_2 < 1$ such that for $\delta \defeq 1/q^{c_1}$, $l \defeq q^{c_2}$, if $f : \F_q^m \to \q$ passes the Low Degree Test (Definition \ref{low-degree-def}) with probability $p$, then there are $l$ degree $d$ polynomials $f^1,f^2,\ldots, f^l$ such that :

$$\Pr_{L, x} \[g_L(x) = f(x) \ \ \& \ \  \exists \ j \in \{1,2,\ldots, l\} \ : \ g_L \equiv f^j|_L  \] \geq p - \delta.$$

In words, whenever the low degree test accepts, except with probability $\delta$, the test picks a line $L$ such that $g_L$ corresponds to the restriction of one of the polynomials $f^1,f^2,\ldots, f^l$ to $L$.

We assume here that $d \ll q$ (in our application, $d \leq O(\log q$)).

\end{theorem}

\subsection{Sum Check Protocol}\label{sec:sum-check}
We will also need the sum check protocol for our PCP. We start with some definition, state the test and the main theorem establishing the soundness of it. 
Think of $M=2m$ and, hence, $\F_{q}^M = \F_{q}^{m} \times \F_{q}^{m}$ in the discussion below. Also one can think of $d=4m.$ We first need a notion of partial sums of polynomials.
\begin{definition}[{\bf Partial Sums}] \label{partial-sum-def}
Let $g:  \mathbb{F}_q^{M} \mapsto  \mathbb{F}_q $ be a degree $d$ polynomial.  For every $0 \leq j \leq M-1$ and every $a_1,a_2, \ldots, a_j \in \mathbb{F}_q$ we define the partial sum $g_{a_1,a_2,\ldots,a_j}$ as a polynomial from $\F_{q} \mapsto \F_{q}$ as follows:

$$ g_{a_1,a_2,\ldots,a_j} (z) \defeq  \displaystyle\sum_{b_{j+2},\ldots, b_{M} \in \{0,1\}} g(a_1,a_2,\ldots,a_j,z,b_{j+2},\ldots,b_{M}). $$

When $j=0$ we denote the polynomial as $g_{\emptyset}$.  When $j=M-1,$ the summation is just $g(a_{1},\ldots, a_{M-1},z).$ Note that all the polynomials so defined are of degree at most $d$.

\end{definition}

\begin{definition}[{\bf Sum Check Protocol}] \label{sum-check-def}
The Sum Check Protocol takes as input a value table for a function $g: \F_q^M \to \q$, a target sum $c \in \F_q$ and for every $0 \leq j \leq M-1$ and every $a_1,a_2, \ldots, a_j \in \mathbb{F}_q$, the coefficients of a degree $d$ polynomial $p_{a_1,a_2,\ldots,a_j}$.  The goal is to check whether $ \displaystyle\sum_{z \in \{0,1\}^M}  g(z)  = c.$  The intention is that $g$ is a degree $d$ polynomial and $p_{a_1,a_2,\ldots,a_j}$ correspond to partial sums of $g$ as in Definition \ref{partial-sum-def}.
The test proceeds by picking $x \defeq (a_1,a_2,\ldots,a_M) \in \mathbb{F}_q^M$ uniformly at random and accepts if and only if all of the following tests pass.

\begin{enumerate}
\item \label{sum-def} $p_{\emptyset}(0) + p_{\emptyset}(1) = c.$

\item \label{consistency-def} For all $1 \leq j \leq M-1$, 
$ p_{a_1,a_2,\ldots,a_{j-1}} (a_j)  = p_{a_1,a_2,\ldots,a_j} (0) + p_{a_1,a_2,\ldots,a_j} (1).$

 \item \label{point-check} $p_{a_1,a_2,\ldots,a_{M-1}} (a_{M}) =  g(x).$

\end{enumerate}

\end{definition}

The following theorem will be used in our reduction. The proof appears in Section \ref{sec:sum-check-proof}.

\begin{theorem}[{\bf Soundness of Sum Check Protocol}] \cite{LFKN92} \label{sum-check-soundness}
Let $g^1, g^2, \ldots, g^l : \F_q^M \to \q$ be degree $d$ polynomials and $g : \F_q^M \to \q$ an arbitrary function.  Suppose for every $1 \leq j \leq l$, $ \displaystyle\sum_{z \in \{0,1\}^M}  g^j(z)  \neq c.$
For $x \in \F_q^M$, let $\mathcal{P}(x)$ be the event that the Sum Check Protocol (Definition \ref{sum-check-def}) accepts on inputs $g$, $c$ and $p_{a_1,a_2,\ldots,a_j}$.  Here $x$ is the choice of randomness in the Sum Check Protocol.  

Then

$$ \Pr_{x \in \F_q^M} \[ \mathcal{P}(x) \ \ \& \ \ \exists \ j \in \{ 1, \ldots, l \} \ : \ g(x) = g^j(x) \] \leq Mdl/q $$

In words, the probability that the Sum Check Protocol accepts when $g$ is consistent with one of $g^1, g^2, \ldots, g^l$ is at most $Mdl/q$ where $g^1, g^2, \ldots, g^l$ are degree $d$ polynomials whose sum is not the required value.

\end{theorem}

\section{The Reduction} \label{reduction}

The following is the main theorem  about the reduction and  implies Theorem \ref{thm:main-cvpp} via Theorem \ref{cvpp}.
\begin{theorem}\label{thm:main}
Unless  $NP \subseteq DTIME(2^{\log^{O(1/\eps)} n})$, {\sc MWSPP} is hard to approximate to factor $2^{{\log^{1-\epsilon}} n}$ for an arbitrary small constant $\epsilon > 0$.
\end{theorem}

Towards the proof of this theorem, we will give a reduction from \fq to {\sc MWSPP}.  The reduction proceeds in three steps:

\begin{itemize}

\item Reduction from \fq to a PCP with low query complexity (Section \ref{sum-check}).

\item Construction of an {\sc HLCPP} instance from the PCP (Section \ref{graph}).

\item Reduction from {\sc HLCPP} to {\sc MWSPP}  (Section \ref{mwspp}).

\end{itemize}
Finally, we will complete the proof in  Section \ref{sec:proof-main} where the choice of parameters is made.

\subsection{Smooth PCP with Low Query Complexity}\label{sum-check}

Note that the \fq instance given by Lemma \ref{sound} has almost all the variables appearing in every equation.  For the reduction to {\sc MWSPP} we require a PCP where every query depends on a few variables.  We will also crucially need a {\em smoothness} property from the PCP similar to the one described for {\sc LCPP} in Section \ref{overview}.  To this end, we use the Low Degree Test of \cite{AS03} and the Sum Check Protocol of \cite{LFKN92}, similarly as in \cite{KP06}. 

\subsubsection{Describing the PCP} \label{pcp}
Let $P$ be the instance of \fq given by Lemma \ref{sound} over variables $\{z_1,\ldots,z_n\}$. 
Let $m \defeq \log n.$ Here we assume that $n$ is a power of $2.$ We think of the variables of $P$ as being embedded into $\{0,1\}^m$ within $\mathbb{F}_q^m$.
Henceforth, we will refer to the variables by their corresponding points in $\{0,1\}^m$.  Thus, an assignment $A : \{0,1\}^m \to \F_q$ to the variables can be extended to a degree $m$ polynomial $f : \mathbb{F}_q^m \mapsto \mathbb{F}_q$ such that $f$ is consistent with $A$ on $\{0,1\}^m$.  

Let the equations be $E_{1},\ldots, E_{q}$ where each equation is of the form
$$ E_{i} \equiv  P_{i}(z_{1},\ldots,z_{n}) = C_{i} \equiv \sum_{s,t \in [n]} c_{i}({s,t}) z_{s}z_{t} =  C_{i} \equiv \sum_{\alpha,\beta \in \{0,1\}^{m}} c_{i}({\alpha,\beta}) z_{\alpha} z_{\beta}= C_{i}.$$
For an assignment $A$ to $\{z_\alpha\}_{\alpha \in \{0,1\}^{m}},$ let $f_{A}$ denote the degree $m$ polynomial encoding $A.$
Now, checking whether an equation $E_i \in P$ is satisfied by $A$ amounts to checking 
$$
 \sum_{\alpha,\beta \in \{0,1\}^{m}} c_{i}({\alpha,\beta}) f_{A}(\alpha) f_{A}(\beta)= C_{i}.
$$
Note that $c_{i}({\alpha,\beta})$ can be thought of as a degree $2m$ polynomial over $\mathbb{F}_q^{2m}$ and is a part of the pre-processing.

The PCP we will construct expects the following tables:

\begin{enumerate}

\item {\bf Points Table:} The value of a function $f : \mathbb{F}_q^{m} \mapsto \mathbb{F}_q$ at every point in $\mathbb{F}_q^{m}$. The intention is that $f$ is a degree $m$ polynomial which encodes a satisfying assignment to $P$ within $\{0,1\}^m$, i.e., for a satisfying assignment $A,$ $f(\alpha)=f_{A}(\alpha)$ for all $\alpha \in \{0,1\}^{m}.$ The size of this table is $q^{m}.$

\item {\bf Lines Table:}  The coefficients of a degree $m$ polynomial $g_L$ for every (affine) line $L$ of $\mathbb{F}_q^{m}$.  The intention is that $g_L$ is the restriction of $f$ on $L$. The size of this table is at most $q^{2m} \cdot (m+1).$

\item {\bf Partial Sums Table:}  For every equation $E_i \in P$, every $0 \leq j \leq 2m-1$ and every $a_1,a_2, \ldots, a_j \in \mathbb{F}_q$, the coefficients of a degree $4m$ polynomial $p_{i,a_1,a_2,\ldots,a_j}$.  The intention is that $p_{i,a_1,a_2,\ldots,a_j}$ correspond to partial sums of $g_i$ (Definition \ref{partial-sum-def}) where $g_i (\alpha,\beta) \defeq c_{i}({\alpha,\beta}) f(\alpha) f(\beta)$ where $\alpha \defeq (a_{1},\ldots, a_{m})$ and $\beta \defeq (a_{m+1},\ldots,a_{2m}).$   Note that $g_i$ has degree at most $4m$ and the size of the $j$-th partial sum table is $q \cdot q^{j} \cdot (4m+1).$

\end{enumerate}

{\bf PCP Test:}

Pick equation $E_i \in P$ uniformly at random.  Pick $\alpha \defeq (a_1,a_2,\ldots,a_m) \in \mathbb{F}_q^m$, $\beta \defeq (a_{m+1},a_{m+2},\ldots,a_{2m}) \in \mathbb{F}_q^m$ uniformly at random.  Let $L$ be the line passing through $\alpha$ and $\beta$.   Read the following values from the corresponding tables:

\begin{itemize}

\item $f(\alpha),f(\beta) \in \mathbb{F}_q$ from the Points table. 

\item The polynomial $g_L$ from the Lines table.

\item The polynomials $p_{i,a_1,a_2,\ldots,a_j}$ from the Partial Sums table for every $0 \leq j \leq 2m-1$.

\end{itemize}

{\bf Acceptance Criteria for the Test:}

Accept if and only if all of the following tests pass.

\begin{enumerate}

\item \label{low-degree} $g_L(\alpha) = f(\alpha)$ and $g_L(\beta) = f(\beta).$  

\item \label{sum} $p_{i,\emptyset}(0) + p_{i,\emptyset}(1) = C_i.$

\item \label{consistency} For all $1 \leq j \leq 2m-1$, 
$ p_{i,a_1,a_2,\ldots,a_{j-1}} (a_j)  = p_{i,a_1,a_2,\ldots,a_j} (0) + p_{i,a_1,a_2,\ldots,a_j} (1).$

 \item \label{nonlinear} $p_{i,a_1,a_2,\ldots,a_{2m-1}} (a_{2m}) =  c_{i}({\alpha,\beta}) f(\alpha) f(\beta).$

\end{enumerate}

Note that we allow arbitrary pre-processing on everything except  $\left\{ C_i \right\}_{i=1}^m$.


\subsubsection{Completeness and Soundness of the PCP}
We prove the following theorem here:

\begin{theorem}[Low Degree and Sum Check]\label{thm:pcp}
Let $P$ be a $\F_{q}$-{\sc QCSPP} instance. Then
\begin{enumerate}
\item If $OPT(P)=1,$ then the PCP Test succeeds with probability $1.$

\item If $OPT(P) \leq k/q$ and $k < q^{c}$ for a small enough $c,$ then the test succeeds above with  probability at most $1/q^{e}$ for some constant $e>0.$

\end{enumerate}
\end{theorem}
The proof of the theorem follows from the following two lemmas.

\begin{lemma}[Completeness]\label{lem:pcp-complete} If there exists an assignment $A$ to $\{z_{1},\ldots, z_{n}\}$ such that $OPT(P)=1,$ i.e., $E_{1},\ldots, E_{q}$ are all satisfied, then there is an assignment to all the tables such that the test accepts with probability $1.$
\end{lemma}
\begin{proof}

We let $f \defeq f_{A},$ $g_{L} \defeq f_{A}|_{L}$ for all $L,$ and for all $i \in [q],$ $0 \leq j \leq 2m-1,$ and $a_{1},\ldots, a_{j } \in \F_{q},$
$$ p_{i,a_{1},\ldots,a_{j}} \defeq \sum _{b_{j+2},\ldots, b_{2m} \in \{0,1\}}  h_{i}(a_{1},\ldots, a_{j},z,b_{j+2},\ldots,b_{2m}),$$
where $h_{i}(x,y)$ is the polynomial of degree at most $4m$ representing  $c_{i}({x,y})f_{A}(x)f_{A}(y).$  It is clear that the test succeeds with probability $1.$
\end{proof}
 
\begin{lemma}[Soundness] \label{pcp-sound}
There is an absolute constant $e >0$ such that if $OPT(P) \leq k/q$ and $k < q^{c}$ for a small enough $c,$ then the PCP described above has soundness at most $1/q^e$.

\end{lemma}

\begin{proof}

We first observe that Step \ref{low-degree} of the protocol is equivalent to running a low degree test (Definition \ref{low-degree-def}) on $L$ and $\alpha$ with input tables $g_L$ and $f$ respectively.  This is because the choice of $\beta$ is independent of $\alpha$ and uniform in $\F_q^m$.  Let $0 < c_1, c_2 < 1$ be the constants given by Theorem \ref{as}.   Let $f^1,f^2,\ldots,f^l$ be the list of $l \defeq q^{c_2}$ polynomials promised by Theorem \ref{as}.  

The following events can happen on a run of the PCP:

\begin{enumerate}

\item The low degree test between $L$ and $\alpha$ fails.  That is, $g_L(\alpha) \neq f(\alpha)$.  In this case, the PCP does not accept.

\item $g_L(\beta) \neq f(\beta)$.  In this case, the PCP does not accept.

\item The low degree test accepts ($g_L(\alpha) = f(\alpha)$) but there is no $1 \leq i \leq l$ such  that  $g_L \equiv f^i |_L$.  By theorem \ref{as}, this happens with probability at most $\delta \defeq 1/q^{c_1}$.

\end{enumerate}

If none of the events listed above occur, then we have that $g_L$ is the restriction of $f_j$ for some $1 \leq j \leq l$.
Also, since Step \ref{low-degree} accepts, we must have $f(\alpha) = f^j(\alpha)$ and $f(\beta) = f^j(\beta)$.

Let $E_i$ be an equation not satisfied by any $f^j$ for $1 \leq j \leq l$.  Note that Steps \ref{sum}, \ref{consistency} and \ref{nonlinear} are equivalent to running the Sum Check Protocol (Definition \ref{sum-check-def}) on $g_i : \F_q^{2m} \to \q$ defined as $g_i(\alpha,\beta) \defeq c_{i}({\alpha,\beta}) f(\alpha) f(\beta)$.  $g_i$ has degree at most $4m$.  Let $g^j_i(\alpha,\beta) \defeq c_{i}({\alpha,\beta}) f^j(\alpha) f^j(\beta)$.  Finally, for $x \in \F_q^{2m}$, let $\mathcal{P}_i(x)$ be the event that the Sum Check Protocol accepts.

Applying Theorem \ref{sum-check-soundness}, 

$$ \Pr_{x \in \F_q^{2m}} \[ \mathcal{P}_i(x) \ \ \& \ \ \exists \ j \in \{ 1, \ldots, l \} \ : \ g_i(x) = g^j_i(x) \] \leq (2m)dl/q $$

Thus, when none of the events in the list occur, the PCP accepts with probability at most $(2m) \cdot (4m) \cdot l/q$ conditioned on choosing $E_i$.  Note that every $f^j$ may satisfy at most $k$ of the $q$ equations.

Thus, the total probability that the PCP accepts is at most $\delta + (1 - lk/q) \cdot O(m^2l/q)$ and it is easy to check that by our choice of parameters this is smaller than $1/q^e$ for some absolute constant $e > 0$.
\end{proof}

\subsection{PCP as Hyper-graph Label Cover} \label{graph}

It will be useful to think of the PCP as a graph labeling problem.  The labeling problem we consider is similar to the well-known {\sc Label Cover} problem except for the following differences:

\begin{itemize}

\item The graph is not bipartite but consists of several layers, with edges between consecutive layers.  In addition, there are hyper-edges which consist of several edges.  The goal is to find a labeling which satisfies the maximum fraction of hyper-edges, where the constraint corresponding to a hyper-edge is the logical AND of the constraint corresponding to each of its edges.

\item The constraints corresponding to edges are not projection constraints as in the case of {\sc Label Cover}, but the more general \emph{many-to-many} constraints.  For an edge $e=(u,v)$, a many-to-many constraint is described by an ordered partition of the label set of $u$ and the label set of $v$ such that the constraint is satisfied if and only if both $u$ and $v$ receive labels from matching partitions.  Formally, let $e=(u,v)$ be an edge and $[R_u],$ $[R_v]$ be the label sets of vertices $u$ and $v$.  Then the many-to-many constraint is described by a pair of maps $\pi_e : [R_u] \to [R_e],$ $\sigma_e : [R_v] \to [R_e]$ where $[R_e]$ is a label set associated to $e$.  A label $l$ to $u$ and a label $l'$ to $v$ is said to satisfy edge $e$ if $\pi_e(l) = \sigma_e(l')$.

\end{itemize}

We now formally describe the {\sc Hyper-graph Label Cover} problem.  While the term {\sc Hyper-graph Label Cover} can be potentially used for a more general class of problems, in this paper we restrict our attention to a very special class of graphs useful for our reduction.

\begin{definition} {\sc Hyper-graph Label Cover Problem}  \label{hyplc}

An instance $G(V, E, \mathcal{E}, [R_0], [R_1], \ldots, [R_{2m+1}], \{\pi_e, \sigma_e\}_{e \in E}, \{\mathcal{S}_v, S_v\}_{v \in \mathcal{L}_0})$ of {\sc Hyper-graph Label Cover} consists of:

\begin{itemize}

\item A graph $G(V,E)$.  The vertices are partitioned into $2m+2$ disjoint layers, $V \defeq \mathcal{L}_0 \cup \mathcal{L}_1 \cup \cdots \cup \mathcal{L}_{2m+1}$.  The edges in $E$ are always between a vertex in $\mathcal{L}_i$ and a vertex in $\mathcal{L}_{i+1}$ for some $i$.

\item Label sets $[R_i]$ for vertices in layer $\mathcal{L}_i$.  Furthermore, for every vertex $v \in \ml_0$, there is a partition $\mathcal{S}_v$ of $[R_0]$ and an {\bf allowable set} of labels $S_v \in \mathcal{S}_v$.

\item A many-to-many constraint for every edge.  Let $e=(u,v)$ be an edge where $u \in \mathcal{L}_i$, $v \in \mathcal{L}_{i+1}$.  The instance contains projections $\pi_e : [R_i] \to [R_e]$, $\sigma_e : [R_{i+1}] \to [R_e]$.  A labeling $(l,l')$ to $(u,v)$ is said to satisfy $e$ if $\pi_e(l) = \sigma_e(l')$.

\item A set of hyper-edges $\mathcal{E}$.  Every hyper-edge consists of one vertex from the first $2m+1$ layers and two vertices from the last layer, such that there is an edge between any pair of vertices in adjacent layers.  A labeling to the graph satisfies a hyper-edge if all the edges contained in it are satisfied.

\end{itemize}

The goal is to find a labeling to the vertices which satisfies the maximum fraction of hyper-edges.  Vertices in $\mathcal{L}_i$ are required to receive a label from $[R_i]$.  Furthermore, vertices in $\mathcal{L}_0$ are required to receive labels from their {\bf allowable set}. 

\end{definition}

We also define a pre-processing version of the {\sc Hyper-graph Label Cover Problem} similar to the LCPP problem of \cite{AKKV05}.

\begin{definition} {\sc Hyper-graph Label Cover Problem with Pre-processing (HLCPP)}  \label{hlcpp}

Given an instance $G(V, E, \mathcal{E}, [R_0], [R_1], \ldots, [R_{2m+1}], \{\pi_e, \sigma_e\}_{e \in E}, \{\mathcal{S}_v, S_v\}_{v \in \mathcal{L}_0})$ of {\sc Hyper-graph Label Cover}, the {\sc HLCPP} problem allows arbitrary pre-processing on all parts of the input except the allowable sets $\{S_v\}_{v \in \mathcal{L}_0}$.

\end{definition}

We will need a notion of smoothness similar to the definition of {\sc Smooth Label Cover}.  

\begin{definition} (Smoothness) \label{def-smoothness}

We say that an {\sc HLCPP} instance $G(V, E, \mathcal{E}, [R_0], [R_1], \ldots, [R_{2m+1}], \{\pi_e, \sigma_e\}_{e \in E}, \{\mathcal{S}_v, S_v\}_{v \in \mathcal{L}_0})$ is $\delta$-smooth if for every $0 \leq i \leq 2m$, $u \in \mathcal{L}_i$, $l \neq l' \in [R_i]$ we have

$$ \Pr_{e = (u,v) \in E} \[ \pi_e(l) = \pi_e(l') \] \leq \delta $$

Here $v \in \mathcal{L}_{i+1}$ and $(\pi_e, \sigma_e)$ is the many-to-many constraint associated to $e$.

\end{definition}

Lastly, we will need that the hyper-edges of the graph are regular in a certain sense.

\begin{definition} (Uniformity) \label{def-uniformity}

Let $G(V, E, \mathcal{E}, [R_0], [R_1], \ldots, [R_{2m+1}], \{\pi_e, \sigma_e\}_{e \in E}, \{\mathcal{S}_v, S_v\}_{v \in \mathcal{L}_0})$ be an {\sc HLCPP} instance.  We say that the instance is uniform if the following conditions are satisfied:

\begin{enumerate}

\item \label{one} For every $0 \leq i \leq 2m+1$, every vertex in layer $\ml_i$ has the same number of hyper-edges passing through it.

\item \label{two} For every $0 \leq i \leq 2m$, the following two distributions are equivalent:

\begin{itemize}

\item Select an edge between a vertex in layer $\ml_i$ and a vertex in layer $\ml_{i+1}$ uniformly at random.

\item Select a hyper-edge $\mathcal{H} \in \mathcal{E}$ uniformly at random and then select an edge from $\mathcal{H}$ between a vertex in layer $\ml_i$ and a vertex in layer $\ml_{i+1}$ uniformly at random.  Recall that a hyper-edge contains exactly one edge between layers $\ml_i$ and $\ml_{i+1}$ for $0 \leq i \leq 2m-1$ and two edges between layers $\ml_{2m}$ and $\ml_{2m+1}$.

\end{itemize}

\end{enumerate}

\end{definition}

We next briefly describe how the PCP described in Section \ref{sum-check} can be thought of as an {\sc HLCPP} instance.

\begin{itemize}

\item {\bf Layers $\mathcal{L}_{2m}$ and $\mathcal{L}_{2m +1}$: } These are the Lines table and the Points table respectively.  
There is a vertex $L$ in $\mathcal{L}_{2m}$ corresponding to every line in $\mathbb{F}_q^m$.  There is a label to $L$ for every possible univariate degree $m$ polynomial over $\mathbb{F}_q$. Hence, the number of vertices in $\mathcal{L}_{2m}$ is at most $q^{2m}$ and the size of the label set for each vertex is $R_{2m} = q^{m+1}.$   
There is a vertex $\alpha$ in $\mathcal{L}_{2m + 1}$ corresponding to every $\alpha \in \mathbb{F}_q^m$.  There is a label $a$ to $\alpha$ for every possible $a \in \mathbb{F}_q$. Hence, the size of the vertex set in $\mathcal{L}_{2m+1}$ is $q^{m}$ and size of the label set is $R_{2m+1} = q.$

There is an edge between $L$ and $\alpha$ if the point $\alpha$ belongs to the line $L$.  The constraint between the two vertices corresponds to Step \ref{low-degree} of the PCP.

\item {\bf Layers $\mathcal{L}_0$ through $\mathcal{L}_{2m}$:} These are the  Partial Sums table and the Lines table respectively.
For $1 \leq j \leq 2m-1$, there is a vertex corresponding to $(i,a_1,a_2,\ldots,a_j)$ in $\mathcal{L}_j$ for every equation $E_i \in P$ and every $a_1,a_2, \ldots, a_j \in \mathbb{F}_q$.  There is a label to $(i,a_1,a_2,\ldots,a_j)$ for every possible univariate degree $4m$ polynomial over $\mathbb{F}_q$.  
For $j=0$, there is a vertex $(i,\emptyset)$ corresponding to every equation $E_i \in P$.  There is a label to $(i,\emptyset)$ for every univariate degree $4m$ polynomial over $\Fq$.  
Furthermore, there is a partition of the label set into $q$ parts, indexed by $\F_{q}$ as follows:
$$ P_{a} \defeq \{ \text{all polynomials $p$ of degree at most } \; 4m \; \text{over $\F_{q}$ such that}  \;p(0)+p(1)=a\}.$$ 
The {\bf allowable set} of labels for every vertex corresponds to the part that satisfies Step \ref{sum} of the PCP.  Thus, for $0 \leq j \leq 2m-1,$ the size of $\mathcal{L}_{j}$ is $q \cdot q^{j}$ while the size the label set is $R_0 = R_1 = \cdots = R_{2m -1} = q^{4m+1}.$ 

For $1 \leq j \leq 2m-1$,  there is an edge between a vertex $(i,a_1,a_2,\ldots,a_{j-1})$ in $\mathcal{L}_{j-1}$ and a vertex $(i',a'_1,a'_2,\ldots,a'_{j})$ in $\mathcal{L}_{j}$ if $i = i'$ and $a_k = a'_k$ for $1 \leq k \leq j-1$.  The corresponding constraints are given by Step \ref{consistency} of the PCP.  

There is an edge between vertex $(i,a_1,a_2,\ldots,a_{2m-1})$ in $\mathcal{L}_{2m-1}$ and vertex $L$ in $\mathcal{L}_{2m}$ if there is an $a_m \in \Fq$ such that for $\alpha \defeq  (a_1,\ldots,a_m)$, $\beta \defeq (a_{m+1},\ldots,a_{2m})$, the line $L$ passes through $\alpha$ and $\beta$.  The corresponding constraints are given by Step \ref{nonlinear} of the PCP.  Note that Step \ref{nonlinear} requires the values of the function at points $\alpha$ and $\beta$ both of which lie on line $L$.  Thus, a label to $L$ specifies the values of $f$ at $\alpha$ and $\beta$.

\end{itemize}

It can be checked that the constraints so defined are many-to-many constraints.\footnote{Actually, the constraint between vertices in layers $\mathcal{L}_{2m-1}$ and $\mathcal{L}_{2m}$ is not many-to-many when $c_i(\alpha,\beta) = 0$ but this happens for at most $2m/q$ fraction of vertices for every equation hence we can afford to ignore these vertices and any hyper-edges containing them.}  Note that we allow pre-processing on everything except the {\bf allowable set} of labels for vertices in layer $\mathcal{L}_0$ as required.

It can be seen that the {\sc HLCPP} instance so constructed is $4m/q$-smooth, since no two distinct degree $4m$ polynomials over $\F_q$ can agree on more than $4m/q$ fraction of points in $\F_q$.

We record this identification of the PCP with an {\sc HLCPP} instance as the following theorem.

\begin{theorem} \label{hlcpp-soundness}

There is a reduction from an \fq instance $P$ over $n$ variables to an {\sc HLCPP} instance $L= G(V, E, \mathcal{E}, [R_0], [R_1], \ldots, [R_{2m+1}], \{\pi_e, \sigma_e\}_{e \in E}, \{\mathcal{S}_v, S_v\}_{v \in \mathcal{L}_0})$ where $m= \log n$, such that 

\begin{enumerate}
\item If $OPT(P)=1,$ then $OPT(L)= 1.$

\item If $OPT(P) \leq k/q$ and $k < q^{c}$ for a small enough $c,$ then $OPT(L) \leq 1/q^{e}$ for some constant $e>0.$

\end{enumerate}

Furthermore, the {\sc HLCPP} instance $L$ is $(4 m/q)$-smooth (Definition \ref{def-smoothness}) and uniform (Definition \ref{def-uniformity}).

\end{theorem}

\subsection{Reduction to {\sc MWSPP}} \label{mwspp}

The reduction from {\sc HLCPP} to {\sc MWSPP} is very similar to the reduction from {\sc LCPP} to {\sc MWSPP} described in \cite{AKKV05}.

Let $G(V, E, \mathcal{E}, [R_0], [R_1], \ldots, [R_{2m+1}], \{\pi_e, \sigma_e\}_{e \in E}, \{\mathcal{S}_v, S_v\}_{v \in \mathcal{L}_0})$ be an instance of {\sc HLCPP}.
For each vertex $v \in V$ and each label $l$ to $v$ we have a variable $w_{v,l}.$  
We now describe the fixed linear forms ${B}_f$ of the {\sc MWSPP} instance.  Below, $\bigoplus$ denotes addition over $\F_{2}.$

 {\bf Vertex constraints:} 
 \begin{enumerate}
 \item $\forall 1 \leq j \leq 2m+1,$ $\forall v \in \mathcal{L}_{j}, \; $  $\bigoplus_{l \in [R_{j}]} w_{v,l}=1.$
 
  \item $\forall v \in \ml_0,$ $\forall S \in \mathcal{S}_v$,
  
  \begin{equation}\label{allow}
\text{  $\bigoplus_{l \in S} w_{v,l} =1$ if $S=S_v$ and $0$ otherwise. }
  \end{equation}
  Notice that only the r.h.s. depends on the input (which is $S_v$).

 \end{enumerate}

%
%

%

{\bf Edge constraints:}

Let $e=(u,v)$ be an edge where $u \in \mathcal{L}_i$, $v \in \mathcal{L}_{i+1}$.  Let $\pi_e : [R_i] \to [R_e]$, $\sigma_e : [R_{i+1}] \to [R_e]$ be the projections describing the many-to-many constraint associated to $e$.  For every element $a \in [R_e]$ we add the following fixed linear form:

\begin{equation} \label{edge-constraint}
 \displaystyle \bigoplus_{l \in [R_i]: \pi_e(l) = a} w_{u,l} = \bigoplus_{l \in [R_{i+1}]: \sigma_e(l) = a} w_{v, l}.
\end{equation}

We now describe the variable forms ${B}_v$ for the {\sc MWSPP} instance.  Let $q_j$ be the number of vertices in layer $\mathcal{L}_j$.  Let $\tilde{q} \defeq \prod_{j=0}^{2m+1} q_j$.  For every layer $\mathcal{L}_j$, $0 \leq j \leq 2m+1$, every vertex $v \in \mathcal{L}_j$ and every label $l$ to $v$, we have the variable form
$ w^{j}_{v,l}  $
repeated $\tilde{q}/q_j$ times. This completes the description of the {\sc MWSPP} instance. It remains to prove the completeness and the soundness of this reduction which we do next.

\subsubsection{Soundness of the {\sc MWSPP} instance}

Here we show that the {\sc MWSPP} instance constructed has a large gap.

\begin{theorem}[Reduction from $\F_{q}$-{\sc QCSPP} to {\sc MWSPP}] \label{mw-sound}
Let $h$ be such that $1/(m^{3}h)^{3m} \geq 1/q^{e}$ for large enough $m$ and for some fixed small constant $e.$  
\begin{itemize}
\item {\bf Completeness:} If $P$ is satisfiable then the {\sc MWSPP} instance constructed in Section \ref{mwspp} has a solution of weight at most $(2m + 2) \cdot \tilde{q}$.

\item {\bf Soundness: } If $P$ is such that $OPT(P) \leq k/q$ then the {\sc MWSPP} instance constructed in Section \ref{mwspp} has no solution of weight less than $h \cdot (2m+2) \cdot \tilde{q}.$

\end{itemize}

\end{theorem}

\begin{proof}

{\bf Completeness.}  
If the $\F_{q}$-{\sc QCSPP} instance $P$ is satisfiable then the {\sc HLCPP} instance has a labeling which satisfies all constraints (Theorem \ref{hlcpp-soundness}).  For an {\sc MWSPP} variable $w^{j}_{v,l}$ corresponding to vertex $v$ and label $l$ to $v$, we let $w^{j}_{v,l} = 1$ if $v$ was assigned the label $l$ and $0$ otherwise.  It is easy to see that this satisfies all fixed linear forms and gives a solution of weight 
$$ \sum_{j=1}^{2m+1} \sum_{v \in \mathcal{L}_{j}} 1 \cdot \tilde{q}/q_{j} = \sum_{j=1}^{2m+1} q_{j} \cdot \tilde{q}/q_{j} = (2m + 2) \cdot \tilde{q}.$$

{\bf Soundness.}  In this case we are given that $OPT(P) \leq k/q$ and, hence by Theorem \ref{hlcpp-soundness}, any labeling to the {\sc HLCPP} instance satisfies at most $1/q^{e}$ fraction of the hyper-edges for some small constant $e.$  The number of hyper-edges in the instance  are $|[q] \times F_{q}^{m} \times \F_{q}^{m}|=q^{2m+1}.$
Suppose there is a solution to the {\sc MWSPP} instance of weight $h \cdot (2m+2) \cdot \tilde{q}$ which satisfies all fixed linear forms.  We will give a (randomized) labeling to the {\sc HLCPP} instance which in expectation satisfies more than $1/(m^{3}h)^{3m} \geq 1/q^e$ fraction of the hyper-edges, contradicting Theorem \ref{hlcpp-soundness}.

Let $\{w_{v,l}\}$ be a solution of weight at most $h \cdot (2m+2) \cdot \tilde{q}$. Call a label $l$ for $v$ {\em nonzero} if $w_{v,l} =1.$ (Note that our variables are allowed only $0/1$ values.) We know from our assumption that 
$$\sum_{j=0}^{2m+1} \sum_{v,l} w_{v,l} \cdot \tilde{q}/q_{j} = h \cdot (2m+2) \cdot \tilde{q}.$$
Let $n_{v}^{j}$ denote the number of nonzero variables for the vertex $v$ in the $j$-th layer. Then the above can be written as
$$ \sum_{j=0}^{2m+1} \sum_{v} n_{v}^{j}/q_{j} = h \cdot (2m+2) .$$ Hence, for all $j,$ $\sum_{v} n_{v}^{j}/q_{j} \leq h \cdot (2m+2).$ Hence by Markov's Inequality, for every $j,$ the fraction of $v$ for which $n_{v}^{j} \geq m^{3}h$ is atmost $h \cdot (2m+2) / (m^{3} \cdot h
) \leq 3/m^{2}$ for large enough $m.$
We call a label $i$ for a vertex $v$ \emph{non-zero} if $w_{v,i} =1$.   We remove all vertices from the graph which have more than $r 
\defeq h \cdot m^3$ non-zero labels.  
This removes at most $3/m^2$ fraction of vertices from each layer.  Next, we remove all hyper-edges containing any vertex removed in this step. To bound this number notice that our graph has this property that number of hyper-edges per vertex of layer $j$ is at most $q^{2m+1}/q_{j}$ (by Item \ref{one} of the uniformity property: Definition \ref{def-uniformity}). Since number of  vertices removed per layer is at most $3q_{j}/m^{2},$ the number of hyper-edges removed in layer $j$ is at most $3q^{2m+1}/q_{j}.$ Hence, the number of  hyper-edges removed  overall is at most $3 \cdot (2m+2) q^{2m+1}/m^{2} \leq 9/m \cdot q^{2m+1}$ for large enough $m.$ 
Thus, the total fraction of hyper-edges removed is at most $9/m$ which is negligible. Thus, we have an {\sc HLCPP} instance where every vertex has at most $r$ non-zero labels and we wish to satisfy more than $1/q^e$ fraction of the queries.

{\bf Labeling.}
  We define a randomized labeling for the {\sc HLCPP} instance: randomly assign a nonzero label independently for each vertex. This is possible as the sum (over $\F_{2}$) of the variables corresponding to each $v$ is $1$ and hence not all variables for a vertex can be $0.$

The next claim shows that the expected fraction of hyper-edges satisfied is at least $r^{-3m} = (h \cdot m^3)^{-3m}$ which is larger than $1/q^e$ by our assumption.

\begin{claim}
Conditioned on the hyper-edge not being removed, the expected fraction of hyper-edges satisfied by the randomized labeling defined above is at least $r^{-3m}$ where $r=hm^{3}.$
\end{claim}

\begin{proof}[Proof of Claim]

We first remove all edges $e$ in the graph for which some pair of non-zero labels map to the same label via the constraint associated to $e$.  Formally, let $e=(u,v)$ be an edge, $l \neq l'$ be two non-zero labels for $u$ and $(\pi_e,\sigma_e)$ be the maps describing the many-to-many constraint associated to $e$.  We remove the edge $e$ if $\pi_e(l) = \pi_e(l')$.  Since the instance is $4m/q$-smooth (Definition \ref{def-smoothness}), taking a union bound over all pairs of non-zero labels implies that the fraction of edges removed in the graph is at most $4mr^2/q$.

Next, we remove all hyper-edges containing any edge removed in the previous step.  Using Item \ref{two} of the uniformity property (Definition \ref{def-uniformity}) and a union bound, it can be seen that the total fraction of hyper-edges removed is at most $3m \cdot 4mr^2/q \leq 12 m^2 r^2/q$, which is negligible by our choice of parameters. Thus, we have an {\sc HLCPP} instance where every vertex has at most $r$ non-zero labels and the many-to-many constraint maps all non-zero labels to distinct labels.

For a hyper-edge to be satisfied, its vertex in $\ml_0$ should receive a label from allowable set.  By Equation \eqref{allow}, there is at least one non-zero label from this set.  Thus, with probability at least $1/r$, we pick an allowed label for a hyper-edge.  

For $0 \leq j \leq 2m$, we will show that if we have assigned label $l$ to vertex $u \in \mathcal{L}_j$, then the probability of assigning a consistent label to any of its neighbors in $\mathcal{L}_{j+1}$ is at least $1/r$.  By a consistent label we mean one which satisfies the constraint on the edge.  

Suppose we have picked a label $l$ for a vertex $u \in \ml_j$.  We claim that the left side of Equation \ref{edge-constraint} is $1$, since there is no non-zero label $l'$ for $u$ such that $\pi_e(l) = \pi_e(l')$.  This means that the r.h.s. is also $1$ (since the fixed linear forms are satisfied).  Hence there must be a non-zero label for $v$ which satisfies the constraint associated with the edge $e = (u,v)$, and this label is assigned to $v$ with probability at least $1/r$ (over the random choice of a labeling).  Hence, the constraint between $u$ and $v$ is satisfied with probability at least $1/r.$  

This shows that for a fixed hyper-edge, the probability (over the randomized labeling) it is satisfied is at least $r^{-(2m+3)}$ which is the number of vertices in the hyper-edge.  Thus, the expected fraction of hyper-edges satisfied is at least $r^{-(2m+3)} \geq r^{-3m}$. This completes the proof of the claim.
\end{proof}
Noticing that by our choice of parameters $1/q^{e} < r^{-3m},$ we obtain a contradiction. Hence, this completes the soundness proof and, hence, the theorem. 
\end{proof}

\subsection{Choice of Parameters and the Proof of Main Theorem}\label{sec:proof-main}

\begin{proof}[Proof of Theorem \ref{thm:main}]
Let $Q$ be the \fq instance given by Theorem \ref{np-complete} over $n$ variables and $k=poly(n)$ equations.  We apply Lemma \ref{sound} to get an \fq instance over $n$ variables and $q$ equations where $q \defeq 2^{(\log n)^{(4/\epsilon)}}$.  We then apply the series of reductions described in Sections \ref{sum-check}, \ref{graph} and \ref{mwspp}.

Let $N$ be the size of the {\sc MWSPP} instance constructed in Section \ref{mwspp}.  It can be checked that $N \leq q^{100m}$, where $m \defeq \log n$ for large enough $m.$  Hence, $N \leq q^{\log ^{2}n}$ for large enough $n.$   We need $m$ and $h$ to satisfy

$\nfrac{1}{r^{3m}}= \nfrac{1}{r^{3 \log n}}= \nfrac{1}{(m^{3}h)^{3m}} \geq \nfrac{1}{q^{e}}.$
This is true if $ \log h \leq \nfrac{\log q}{\log ^{2}n}$ and $\log q \gg \log n \log \log n$ and $n$ is let to be large enough.
%


We set $h \defeq q^{{\log}^{-2} n}$.  For a large enough positive integer $D=\nfrac{4}{\eps},$ let $q$ be such that $\log q \defeq \log^{D}n.$ Hence, $\log q \gg \log n \log \log n.$ Moreover $\log N \leq \log ^{D+2}n$ and $\log h = \log ^{D-2} n .$ This implies that 
$$\log ^{1-\eps}N  = \log ^{(D+2)(1-\epsilon)} n   \leq  \log ^{D-2} n  = \log h$$
Finally, $N \leq q^{\log ^{2}n} = 2^{\log ^{O(1/\eps)}n}.$ Summarizing, our reduction is deterministic, the hardness factor is $2^{\log ^{1-\eps}N}$ and takes time $2^{\log ^{O(1/\eps)}n}$ and, hence, holds under the hypothesis  $NP \not \subseteq DTIME(n^{\log ^{O(1/\eps)}n}).$ 
\end{proof}

\newpage

\bibliographystyle{alpha}
\bibliography{refs-all}

\newpage

\appendix

\section{Omitted Proofs}\label{sec:basic-proofs}

\subsection{{\sc $\F_q$-QCSPP} is NP-complete} \label{sec:np-complete}

\begin{theorem} 

{\sc $\F_q$-QCSPP} is NP-complete for all $q=2^r$.

\end{theorem}

\begin{proof}

    We reduce {\sc 3SAT} to \fq.  For this proof, it is convenient
    to view the input for {\sc 3SAT} in the following form: the input is $(V,E),$ where $V,E \in
    \{0,1\}^{m \times n}$ and corresponds to a {\sf 3SAT} formula $\phi=C_1
    \wedge \cdots \wedge C_m$ with variables $\{x_1,\ldots,x_n\}.$   Each row of $V$ corresponds to a clause $C_i$ and $V_{ij}$ is $1$ if and only if $x_j$ appears in $C_i$.
     Thus,
    each row of $V$ has exactly three $1$'s.  The entry $E_{ij}$ is $1$ if and only if the variable $x_j$ appears as a negated literal in $C_i$.

    Since {\sc 3SAT} is in {\sc
    NP}, for every $n,$ there is a circuit $\mathcal{C}_n$ which takes as
    input $(V,E)$ and an assignment ${a}\in \{0,1\}^n,$ such that,
    $\mathcal{C}_n(a,V,E)=1$ if ${a}$ is a
    satisfying assignment for $\phi,$ and $0$ otherwise.

    Now we present the reduction, which is exactly the same as in Theorem 4.2 of \cite{AKKV05}, except that we work over $\F_q$ rather than $\F_2$.  Let $(V,E)$ be the input corresponding
    to a {\sc 3SAT} instance $\phi.$ We may assume that every gate in
    $\mathcal{C}_n$ has fan-in 2 and fan-out 1.  For every bit in the input
    $({a},V,E)$ to $\mathcal{C}_n,$ there is a variable in $\F_q$: $x_i$ is
    supposed to be assigned the $i$-th bit of ${a},$ $x_{ij}$ is
    supposed to be assigned $V_{ij},$ while $x'_{ij}$ is supposed to be
    assigned $E_{ij}.$

    Associated to the output of the $i$-th internal gate\footnote{A gate is
    said to be internal if its output is not an output of the circuit.}  in
    $\mathcal{C}_n$ is a variable $z_i.$ Further, let $y_0$ be the variable
    corresponding to the output gate which outputs whether an assignment
    ${a}$ satisfies $\phi$ or not. 
    
The computation of any gate can be written as a quadratic polynomial
(over $\F_2$) in its inputs (call these $z, z'$) and output (call it $z''$): 
$z'' = zz'$ for an AND gate, $z'' = 1+(1+z)(1+z')$ for an OR gate, and 
$z'' = 1+z$ for a NOT gate.  

Note that $\F_2$ is a sub-field of $\F_q$ since $q=2^r$ is a power of $2$.  Thus, each element of $\F_q$ can be naturally identified with a vector in $\F_2^r$ such that addition in $\F_q$ corresponds to vector addition and multiplication in $\F_q$ corresponds to taking dot products.  In any such representation, the element $0 \in \F_q$ corresponds to the all $0$'s vector while the element $1 \in \F_q$ corresponds to the all $1$'s vector.  The crucial observation is that the polynomials for the AND, OR and NOT gates described above act as co-ordinate wise AND, OR and NOT gates when the inputs and the output are elements of $\F_2^r$ ($\F_q$), when all computation is done over $\F_q$.

    We write such an equation for every gate in
    $\mathcal{C}_n.$ Each equation is of degree at-most $2$ and has at-most
    $3$ variables.  Note that every such equation depends only on the
    description of $\mathcal{C}_n.$  
Finally, we add the additional set of equations $y_0=1,$
    $x_{ij}=V_{ij}$ and $x'_{ij}=E_{ij}.$ Hence, we get a \fq instance over the set of variables
    \begin{multline*}
        \{x_i: i \in [n]\} \cup \{x_{ij}: i\in [m], j \in [n]\} \cup
        \{x'_{ij}:  i \in [m],  j \in [n]\} \cup \{z_i: 1\leq i \leq {\rm
        size}(\mathcal{C}_n)\} \cup \{ y_0 \}. 
    \end{multline*}

    Notice that $\mathcal{C}_n$ can be generated by a polynomial time
    algorithm which is given as input $1^n.$ Hence, this reduction is a
    polynomial time reduction.  
    
    We claim that this
    quadratic system has a solution (over $\F_q$) if and only if $\phi$ has a satisfying
    solution.  The corresponding claim when all variables take values in $\F_2$ follows by construction.  Now note that if there is a solution over $\F_q$ then taking the last co-ordinate of the variables (when viewed as vectors over $\F_2^r$) is a valid solution over $\F_2$, since all gates act co-ordinate wise.
    
 The reduction described above gives constraints which are of degree at most $2$, but not homogeneous.  This is easy to fix by introducing an auxiliary variable $z_0$ and adding the constraint $z_0 z_0 = 1$.  We then multiply all terms of degree less than $2$ by $z_0$.
    
    This completes the proof of the lemma.

\end{proof}

\subsection{Boosting  Soundness through Codes}\label{sec:boost-proof}
We first need some basic definitions.
\begin{definition} {\bf Codes:}  A matrix $C \in \F_q^{m \times k}$ is said to be a generator of the linear code $\{Cx : x \in \F_{q}^{k}\}$ with distance $1 - \delta$ if for any $x \neq y \in \F_q^k$, $\ C(x)$ and $C(y)$ agree on at most $\delta m$ co-ordinates.

\end{definition}

\begin{Fact}[{\bf Reed-Muller Codes}]\label{reed-muller}
For any $q$, let $\Fq$ be the field over $q$ elements.  There is a family of linear codes with generator matrix $C_k \in \F_q^{q \times k}$ with distance $1 - k/q$.  These are the so called Reed Muller codes over $\Fq,$ where the message is thought of as the coefficients of a  degree $k$ polynomial and the codeword the evaluation of this polynomial on all the points in $\F_{q}.$

\end{Fact}

\begin{lemma} \label{sound-1}
Let $Q$ be an instance of \fq over $n$ variables and $k = poly(n)$ equations, for any $q=2^r$.  There is an instance $P$ of \fq over the same set of variables and $q$ equations such that:
\begin{itemize}
\item If $OPT(Q) =1$ then $OPT(P) = 1$ and 
\item if $OPT(Q) < 1$ then $OPT(P) \leq k/q.$
\end{itemize}
\end{lemma}

\begin{proof}

Let $R \in \F_q^{q \times k}$ be the Reed-Muller code matrix as in Fact \ref{reed-muller}.  Let $p_1, \ldots, p_k$ be the equations of $Q$ and let $r \in \F_q^k$ be a row of $R$.  We add the constraint $\sum_{i=1}^k r_i p_i$ to $P$ which is a $\Fq$-linear combination of the equations in $Q$.  Thus, $P$ has $q$ equations.

It is clear that any satisfying assignment to all equations of $Q$ is also a satisfying assignment for all equations of $P$.  This shows that if $OPT(Q) =1$ then $OPT(P) = 1$.

On the other hand, suppose $OPT(Q) < 1$ and fix any assignment $A$ to the variables of $Q$.  An equation $e_i \in Q$ is of the form $p_i(z_1,\ldots,z_n) = c_i$.  Let $v^A \in \F_q^k$ be defined as $v^A_i \defeq   p_i(A(z_1),\ldots,A(z_n)) - c_i$.  Since $OPT(Q) < 1$, $v^A \neq 0^k$.  Thus, by the code property we have that $C_k \cdot v^A$ is zero in at most $k$ co-ordinates.  Notice that $C_k \cdot v^A$ has a $0$ in a co-ordinate if and only if the corresponding equation in $P$ is satisfied by $A$.  Since this holds for every assignment $A$, the theorem follows.

\end{proof}

\subsection{Sum Check Protocol}
\label{sec:sum-check-proof}

\begin{theorem}[{\bf Soundness of Sum Check Protocol}] \cite{LFKN92} 
Let $g^1, g^2, \ldots, g^l : \F_q^M \to \q$ be degree $d$ polynomials and $g : \F_q^M \to \q$ an arbitrary function.  Suppose for every $1 \leq j \leq l$, $ \displaystyle\sum_{z \in \{0,1\}^M}  g^j(z)  \neq c.$
For $x \in \F_q^M$, let $\mathcal{P}(x)$ be the event that the Sum Check Protocol (Definition \ref{sum-check-def}) accepts on inputs $g$, $c$ and $p_{a_1,a_2,\ldots,a_j}$.  Here $x$ is the choice of randomness in the Sum Check Protocol.  

Then

$$ \Pr_{x \in \F_q^M} \[ \mathcal{P}(x) \ \ \& \ \ \exists \ j \in \{ 1, \ldots, l \} \ : \ g(x) = g^j(x) \] \leq Mdl/q $$

In words, the probability that the Sum Check Protocol accepts when $g$ is consistent with one of $g^1, g^2, \ldots, g^l$ is at most $Mdl/q$ where $g^1, g^2, \ldots, g^l$ are degree $d$ polynomials whose sum is not the required value.

\end{theorem}

\begin{proof}
We will prove the theorem by induction on $M$.

{\bf Base Case: M=1} 
We consider two cases:

\begin{enumerate}

\item $p_{\emptyset} = g^j_{\emptyset}$ for some $1 \leq j \leq l$.  In this case Step \ref{sum-def} fails by our assumption on $g^j$.

\item $p_{\emptyset} \neq g^j_{\emptyset}$ for all $1 \leq j \leq l$.  In this case, 

$$
\begin{array}{clr}
&  \Pr_{x \in \F_q} \[ \mathcal{P}(x) \ \ \& \ \ \exists \ j \in \{ 1, \ldots, l \} \ : \ g(x) = g^j(x) \]  & \\
\leq & \Pr_{x \in \F_q} \[ g(x) = p_{\emptyset}(x) \ \ \& \ \ \exists \ j \in \{ 1, \ldots, l \} \ : \ g(x) = g^j(x) \] & $(Since Step \ref{point-check} accepts)$ \\
= & \Pr_{x \in \F_q} \[ \exists \ j \in \{ 1, \ldots, l \} \ : \ p_{\emptyset}(x) = g^j(x) \] & \\
\leq & ld/q &
\end{array}
$$

The last inequality uses the fact that any two distinct degree $d$ polynomials can agree on at most $d/q$ fraction of the points followed by a union bound.

\end{enumerate}

{\bf Inductive Case: M = N}  We again consider two cases as before:

\begin{enumerate}

\item $p_{\emptyset} = g^j_{\emptyset}$ for some $1 \leq j \leq l$.  In this case Step \ref{sum-def} fails by our assumption on $g^j$.

\item $p_{\emptyset} \neq g^j_{\emptyset}$ for all $1 \leq j \leq l$.  In this case, the fraction of points $a \in \mathbb{F}_q$ such that 

\begin{equation} \label{patching}
 \displaystyle\sum_{b_{2},\ldots b_{N} \in \{0,1\}} g^j (a,b_2,\ldots , b_{N}) = p_{\emptyset} (a)
\end{equation}

 for some $1 \leq j \leq l$ is at most $ld/q$.

Note that for a fixed $a \in \F_q$, Steps \ref{consistency-def} and \ref{point-check} are equivalent to running the Sum Check Protocol for checking  $$ \displaystyle\sum_{b_{2},\ldots b_{N} \in \{0,1\}} g (a,b_2,\ldots , b_{N}) = c' $$ where $c' \defeq p_{\emptyset}(a)$.  For $x \in \F_q^{N-1}$, let $\mathcal{P}_a(x)$ be the event that this protocol accepts.

If Equation \ref{patching} does not hold for any $1 \leq j \leq l$ then we can use the inductive assumption to get

$$ \Pr_{x \in \F_q^{N-1}} \[ \mathcal{P}_a(x) \ \ \& \ \ \exists \ j \in \{ 1, \ldots, l \} \ : \ g(a, x) = g^j(a, x) \] \leq (N-1)dl/q $$

Thus, the total probability of acceptance is at most $ld/q + (N-1)dl/q \leq Ndl/q$.

\end{enumerate}
\end{proof}

\end{document}